\documentstyle[11pt,aaspp4,epsf,flushrt]{article}
\newcommand{\lsim}{\raisebox{0.3mm}{\em $\, <$} \hspace{-3.3mm}
\raisebox{-1.8mm}{\em $\sim \,$}}
\newcommand{\gsim}{\raisebox{0.3mm}{\em $\, >$} \hspace{-3.3mm}
\raisebox{-1.8mm}{\em $\sim \,$}}

\begin{document}

\title
{Formation of Large-Scale Obscuring Wall and AGN Evolution
Regulated by Circumnuclear Starbursts}

\author{Ken {\sc Ohsuga} and Masayuki {\sc Umemura}}

\affil{Center for Computational Physics, University of Tsukuba,
Tsukuba, Ibaraki 305-8577, Japan}

\begin{center}
To appear in {\it The Astrophysical Journal Letters}
\end{center}

\begin{abstract}
By considering the radiative force by a circumnuclear starburst 
as well as an AGN,
we analyze the equilibrium configuration and the stability of 
dusty gas in the circumnuclear regions.
It is found that the radiative force by an intensive starburst 
can support a stable gaseous wall with a scale-height 
of several hundred parsecs. Moreover, by taking the simple stellar 
evolution in the starburst into account, 
we find that the covering factor of the wall
decreases on a time-scale of several $10^7$ yr. 
The large-scale wall, if formed, works to obscure the nucleus
due to the dust opacity.
Hence, it is anticipated that the index of AGN type tends to
shift from higher to lower in several $10^7$ yr
according as the circumnuclear starburst becomes dimmer.
On the other hand, if the AGN itself is brighter than the
circumnuclear starburst 
(e.g. quasar case), no stable large-scale wall forms. 
In that case, the AGN is highly probably identified as type 1.
The present mechanism 
may provide a physical explanation for 
the putative correlation between AGN type and host properties that
Sy2's are more frequently associated with circumnuclear starbursts 
than Sy1's, whereas quasars are mostly observed as type 1 regardless 
of star-forming activity in the host galaxies.

\end{abstract}


\section{Introduction}

Recently, intriguing evidences regarding the host galaxies
of active galactic nuclei (AGNs) and quasars (QSOs) have been
accumulated.
%
First, it has been reported that host galaxies of Seyferts are intrinsically 
unlike between type 1 (Sy1) and type 2 (Sy2) 
(Heckmann et al. 1989; Maiolino et al. 1995, 1997, 1998, 1999;
P\'erez-Olea \& Colina 1996; Hunt et al. 1997; Malkan et al. 1998; 
Storchi-Bergmann \& Schmitt 1998).
Sy1's inhabit earlier type quiescent hosts, while Sy2's are 
frequently associated with circumnuclear starbursts,
which often lie in barred galaxies.
The observations may indicate that Sy2's are in an earlier evolutionary 
stage than Sy1's (Radovich et al. 1998), 
whereas in the unified model (Antonucci 1993, for a review)
this dichotomy is simply accounted for with
the orientation of the nucleus with an obscuring torus of subparsec scale.
Second,
the recent {\it HST} images of nearby QSOs have
shown that luminous QSO phenomena occur preferentially in
luminous host galaxies, often being ellipticals
(McLeod \& Rieke 1995b; Bahcall et al. 1997; Hooper et al. 1997).
Also, at high redshifts, towards the QSO H1413+117 at $z=2.546$ 
(``cloverleaf'', a gravitationally lensed quasar)
and QSO BR1202-0725 at $z=4.69$, 
a large amount of dust have been detected, i.e., 
$\sim 10^{9} M_\odot$ for  H1413+117 and $\sim 10^{8} M_\odot$ for BR1202-0725
(Barvainis et al. 1992; Omont et al. 1996).
Molecular gas of at least $10^{11} M_\odot$ are also found
for BR1202-0725 (Ohta et al. 1996a).
These suggest that active star formation are on-going around 
the quasars. 
QSOs, however, are mostly identified as type 1 and 
only a few type 2 QSOs are discovered so far
(Almaini et al. 1995; Ohta et al. 1996b; Brandt et al. 1997).
Hence, a circumnuclear starburst does not seem to urge
type 2 as far as quasars are concerned. 
QSOs are distinctive from Seyferts in that the host
galaxy is in general fainter than the AGN itself
(McLeod \& Rieke 1995b; Bahcall et al. 1997; Hooper et al 1997).
These facts on the host properties for Seyferts and QSOs
suggest a possibility that the AGN type has a close relation 
to circumnuclear starburst events 
and the relative luminosity of starburst to the AGN. 

We consider a physical mechanism which may connect circumnuclear 
star-forming activities with AGN type.
Here, attention is concentrated on the radiative force 
by the circumnuclear starburst.
In ultraluminous IR galaxies, the observed IR luminosities 
(Scoville et al. 1986; Soifer et al. 1986)
are comparable to or greater than the Eddington luminosity for dust opacity
(Umemura et al. 1998, 1999).
Hence, the radiative force is very likely to play an important role 
on the circumnuclear structure of $\sim100$ pc.
In this Letter, supposing the mass distribution in circumnuclear regions,
we analyze the equilibrium configuration and the stability of dusty gas 
which is supported by radiative force by a starburst and an AGN.
In addition,
taking the stellar evolution in the starburst regions into
consideration, we investigate the time evolution of gas distributions,
and attempt to relate the luminosity of the circumnuclear 
starburst to the evolution of the AGN type.

\section{Radiatively-Supported Obscuring Wall}

The circumnuclear starburst regions frequently exhibit ring-like 
features and have radial extension of $\sim 10$ pc up to kpc
(Wilson et al. 1991; Forbes et al. 1994; Mauder et al. 1994; 
Buta et al. 1995; Barth et al. 1995; Maoz et al. 1996; 
Leitherer et al. 1996; Storchi-Bergman et al. 1996).
Thus, here we consider a ring of starburst. (If the starburst regions 
are anisotropic or clumpy, the effects are expected to be smeared out as 
discussed later.)
We calculate the radiation force and the gravity which are exerted 
on the dusty gas. Here, the gravitational potential is determined 
by four components, the galactic bulge, the central black hole, 
the gas disk, and the starburst ring.
We assume the galactic bulge to be an uniform sphere whose 
mass and radius are $M_{\rm bul}$ and $R_{\rm bul}$,
the mass of the central black hole to be $M_{\rm BH}$,
the gas disk to be a Mestel disk whose mass and radius 
are $M_{\rm disk}$ and $R_{\rm disk}$,
and the starburst ring to be an uniform torus
whose mass, curvature radius, thickness, and bolometric luminosity are
$M_{\rm SB}$, $R_{\rm SB}$, $a_{\rm SB}$, and $L_{\rm SB}$, respectively 
(see Figure 1).
The observations of IRAS galaxies by Scoville et al. (1991) 
show that the central regions within several hundred parsecs
possess the gas of $\lsim 10^{10} {\rm M}_\odot$.
By taking this fact into account, we adopt the mass ratio as
$M_{\rm bul}:M_{\rm BH}:M_{\rm disk}:M_{\rm SB}=1:0.01:0.1:1$.
Also, it is found that the starburst ring often consists
of compact star clusters of $\lsim 10$ pc
(
Barth et al. 1995; Maoz et al. 1996; Leitherer et al. 1996).
The size of these clusters is about a tenth of radial extension
of the starburst ring. Therefore, we assume
$R_{\rm bul}:R_{\rm disk}:R_{\rm SB}:a_{\rm SB}=10:1:1:0.1$.
In addition, the nucleus is postulated to be a point source whose 
bolometric luminosity is $L_{\rm nuc}$.

Here, the material is assumed to be subject to the radiative 
force directly by the starburst radiation.
The radiation flux by an infinitesimal volume element, $dV$, 
of the starburst ring is given by 
$dF^i(r,z) = ( \rho_{\rm SB}/4\pi\l^2 ) n^i dV$,
at a point of $(r,z)$ in cylindrical coordinates, 
where $i$ denotes $r$ or $z$,
$\rho_{\rm SB}$ ($=L_{\rm SB}/2 \pi^2 a_{\rm SB}^2 R_{\rm SB}$)
is the luminosity density of the starburst ring, 
$l$ is the distance from $(r,z)$ to this element, and 
$n^i$ is a directional cosine. 
Hence,
the radiation flux force by the starburst ring and the nucleus
at $(r,z)$ is given by
\begin{equation}
 f^i_{\rm rad}=\frac{\chi}{c}\int \frac{\rho_{\rm SB}}{4 \pi l^2}
  n^i dV + \frac{\chi}{c} \frac{i L_{\rm nuc}}
  {4\pi \left( r^2+z^2 \right)^{3/2}},
\end{equation}
where $\chi$ is the mass extinction coefficient for the dusty gas 
(Umemura et al. 1998) and $c$ is the light speed.
Using the above equation,
the equilibrium between the radiation force and the gravity
is written as
\begin{equation}
f_{\rm rad}^z+f_{\rm grav}^z=0,
\end{equation}
in the vertical directions and 
\begin{equation}
\frac{j^3}{r^2}+f_{\rm rad}^r+f_{\rm grav}^r=0, \label{eqr}
\end{equation}
in the radial directions,
where $j$ is the specific angular momentum of the dusty gas and
$f_{\rm grav}^i$ is the gravitational force.

In figure 2, in the case that the starburst ring 
is a dominant radiation source (Case A),
the resultant equilibrium branches are shown in the $r$-$z$ space. 
Here, $\Gamma_{\rm SB}$ and $\Gamma_{\rm nuc}$ are the Eddington measures 
defined by
$\Gamma_{\rm SB}=L_{\rm SB}/(4\pi c G M_{\rm total}/ \chi)$ and 
$\Gamma_{\rm nuc}=L_{\rm nuc}/(4\pi c G M_{\rm total}/ \chi)$ respectively
with $M_{\rm total}=M_{\rm bul}+M_{\rm BH}+M_{\rm disk}+M_{\rm SB}$.
The solid and dashed 
curves represent the equilibrium branches which are stable 
in vertical directions. Above the curves,
the vertical component of the gravity which works to lower the gas
is stronger than the radiation force,
while below the curves the radiation force lifts the gas towards 
the curves.
The dotted curves show the vertically unstable branches. 
The gas is accelerated upwards by the radiative force 
above the curves or falls downwards by the gravity  
below the curves.

In order to get the configuration of finally stable equilibrium,
the stability in radial directions must be taken also
into consideration.
On the dashed curves, the effective potential in the radial directions
turns out to be locally maximal. 
Thus, the dashed branches are unstable points of
saddle type. Finally, only solid curves are stable branches.
%
Figure 2 shows that stable branches emerge only for $\Gamma_{\rm SB}\leq 1$.
This agrees with a naive expectation.
If $\Gamma_{\rm SB}> 1$,
the radiation force blows out the dusty gas in most regions.
(e.g. see a dotted curve of $\Gamma_{\rm SB} = 2$).
When $\Gamma_{\rm SB} \leq 1$, the covering factor of 
the stable wall is a function of $\Gamma_{\rm SB}$.
If $\Gamma_{\rm SB} \sim 1$, the wall
surrounds both the nucleus and the starburst ring.
When the starburst luminosity is smaller
than $\Gamma_{\rm SB} = 0.55$, the wall forms 
only in the vicinity of the starburst ring and exhibits 
a torus-like configuration. 
The $A_V$ of the wall is expected to be at least several,
because radiative force directly from a starburst can be exerted  
on such a wall. 
Then, this large-scale wall of dusty gas would work obscure
the nucleus. When the flux force of scattered diffuse radiation 
operates efficiently in the wall, the wall of larger optical depth may be
supported and therefore $A_V$ could be much larger. 
(The detail should be investigated by multi-dimensional radiation 
hydrodynamics, which will be performed in the future analysis.)
If $A_V$ of the wall is only several magnitudes, 
the AGN would be changed to an intermediate 
type between type 1 and 2, e.g. type 1.3, 1.5 and so on, while
the $A_V$ greater than ten magnitudes would result in the perfect
shift from type 1 to type 2. 

Next, we examine the case that the nucleus is brighter than 
the circumnuclear starburst (Case B).
In this case, only vertically stable branches emerge
even if $\Gamma_{\rm SB}+\Gamma_{\rm nuc}<1$ (see Fig. 3).
On the dashed curves in Fig. 3, contrastively to the Case A, 
there is no solution for the radial equilibrium,
regardless of the value of starburst luminosity.
The gas around the dashed curves is swung away due to the cooperation of 
radiative force and angular momentum.
Resultantly, the formation of the stable wall is precluded 
in the Case B.
This implies that the luminous nuclei like QSOs are 
not likely to be obscured, which are therefore 
mostly identified as type 1.


Further, for the Case A, we consider the effects of stellar 
evolution in the starburst regions on the stable equilibrium 
branches. We assume a Salpeter-type 
initial mass function (IMF), $\phi=A(m_*/M_{\odot})^{-1.35}$, 
the mass-luminosity relation, $(l_*/L_{\odot})=(m_*/M_{\odot})^{3.7}$,
and the mass-age relation, 
$\tau=1.1\times10^{10} {\rm yr}(m_*/M_{\odot})^{-2.7}$,
where $m_*$ and $l_*$ are respectively the stellar mass and luminosity.
Recently, it has been revealed that in starburst regions 
the IMF is deficient 
in low-mass stars, with the cutoff of about $2M_{\odot}$,
and the upper mass limit is inferred to be around $40M_{\odot}$
(Doyon et al. 1992; Charlot et al. 1993;
Doane \& Mathews 1993; Hill et al. 1994; Brandl et al. 1996).
Using the IMF for a mass range of $[ 2M_\odot,40M_\odot ]$ 
and the above relations,
the total stellar luminosity of starburst regions 
is given by a function of time as
$
L_*=
1.5\times 10^{10} \left( 87 t_7^{-0.87}-1\right) 
\left(M_{SB}/10^{10}M_{\odot} \right)L_{\odot}
$,
where $t_7$ is the elapsed time after the coeval starburst
in units of $10^7 \rm yr$.
Also, if we postulate that stars of $>8M_\odot$ are destined to 
undergo supernova explosions and release the energy radiatively
with the efficiency of $\varepsilon$ to the rest mass energy,
the total supernova luminosity is 
$
L_{\rm SN}= 
1.7\times 10^{11} t_7^{-0.87}
\left( {M_{\rm SB}/10^{10}M_{\odot}} \right) 
\left( \varepsilon/10^{-4} \right) L_{\odot}
$
until $t_7=4.0$ and 
$L_{\rm SN}=0$
when $t_7>4.0$.
Hence, the total luminosity of the starburst ring is given by
\begin{equation}
 L_{\rm SB}(t_7)=L_*+L_{\rm SN}.
\end{equation}
Using this dependence on time, 
the luminosity can be translated into the age of the starburst regions.
Therefore, the values of $\Gamma_{\rm SB}$ in figure 2
represent the evolutionary stage of the circumnuclear starburst.
For instance, if we adopt $M_{\rm SB}=10^{10} M_\odot$,
$\Gamma_{\rm SB}=1$ and $0.55$ correspond to 
$4.2\times 10^7 \rm yr$ and $8.1\times 10^7\rm yr$, respectively.

To summarize,
if $\Gamma_{\rm SB}>1$ in the early evolutionary stage,
the dusty gas is blown away by radiative acceleration.
Since the blown-out dusty gas would emit the strong IR radiation,
we may recognize the objects as ultraluminous infrared galaxies.
When $\Gamma_{\rm SB}$ becomes just below unity, both the nucleus 
and the starburst ring are surrounded by the dusty wall.
Then, the AGN is likely to be type 2.
In the later stages, the dusty gas forms a torus-like 
obscuring wall, which shrinks on a time-scale of 
several $10^7$ yr. Then, the AGN tends to be identified as type 1
for a wide viewing angle.
This implies that the type of AGN evolves from higher to lower 
in several $10^7$ yr according as
the circumnuclear starburst becomes dimmer.

\section{Discussion}

Here, we have assumed that the nuclear activity and
the circumnuclear starburst are the simultaneous events.
A solution, for instance, which links the two events,
is the radiatively-driven mass accretion onto a central black hole
due to the radiation drag
(Umemura et al. 1997, 1998; Ohsuga et al. 1999).
However, in very early luminous phases of the starburst, 
the mass accretion onto the black hole is prevented due to the super-Eddington 
radiative force. It result in a radiative blizzard in nuclear regions.
Thus, the nucleus is just identified as an ultraluminous infrared 
galaxy without being accompanied by an AGN.
We predict in the present model that ultraluminous infrared 
galaxies evolve into Seyferts or QSOs in later less luminous phases 
of the starburst.

The circumnuclear starburst could not be axisymmetric but clumpy.
However, the rotational time scale of the starburst ring is shorter than
the shrinking time-scale of the obscuring wall. In the present case, 
the former is around $3.0 \times 10^6 {\rm yr}$ and the latter is
several $10^7 {\rm yr}$. Therefore, the anisotropies of the starburst 
are expected to be smeared out by a 'wheel effect'.
The stable obscuring wall might be subject to the other local
instabilities, i.e., Rayleigh-Taylor or self-gravitational instabilities.
The density gradient of the dusty wall is positive
inside the equilibrium surface and negative outside the surface.
They are in the same directions as the effective acceleration.
Thus, the wall would not be subject to Rayleigh-Taylor instabilities.
As for the self-gravitational instability,
the time-scale of the instability could be as short as $\lsim 10^6$ yr.
So, the wall may fragment on a time-scale shorter than 
the evolutionary time-scale of the wall. Then,
numerous compact gas clouds would form in the wall.
They would emit the narrow emission lines because
they have the velocity dispersion of several 100 km s$^{-1}$.
Also, if the compact clouds are optically thick, 
the radiative force is less effective for them,
so that they fall into the central regions.
They also may partially obscure the nucleus.

In this letter, 
we do not argue that a conventional obscuring torus of 
subparsec scale is dispensable.
Even if the inner obscuring torus may operate to
intrinsically differentiate the type of AGNs, 
the present large-scale wall can work also to raise 
further the type index.
In particular, 
the present mechanism may provide a physical solution 
to account for the tendency 
that Sy2's are more frequently associated with circumnuclear starbursts 
than Sy1's, whereas quasars are mostly observed as type 1 regardless 
of star-forming activity in the host galaxies.
If we adopt the size and mass that conform to realistic values, 
e.g., $R_{\rm SB}\sim$ 100 pc and $M_{\rm SB} \sim 10^{10} M_\odot$,
the obscuring wall is extended to several 100 pc.
Interestingly, it is recently reported that 
the spectra of a sample of AGNs are more consistent with 
obscuring material extended up to $\gsim$ 100 pc around the nuclei
(Rudy et al. 1988; Miller et al. 1991; Scarrott et al. 1991; Goodrich 1995;
McLeod \& Rieke 1995a; Maiolino et al. 1995;
Maiolino \& Rieke 1995) and the hosts of Sy'2 possess more frequently
extended dust lanes (Malkan et al. 1998).
Also, the covering factor of a dusty torus around a QSO, MG 0414+0534, 
is fairly small (Oya et al. 1999). 
These observations are quite intriguing in the light of the present picture. 

\acknowledgments

We are grateful to T. Nakamoto, H. Susa, and S. Oya for helpful
discussion.
The calculations were carried out at the Center for Computational Physics 
in University of Tsukuba. This work is 
supported in part by Research Fellowships of the Japan Society
for the Promotion of Science for Young Scientists, 6957 (KO)
and the Grants-in Aid of the
Ministry of Education, Science, Culture, and Sport, 09874055 (MU). 

\newpage
\begin{figure}
  \centerline{\epsfxsize=400pt \epsfbox{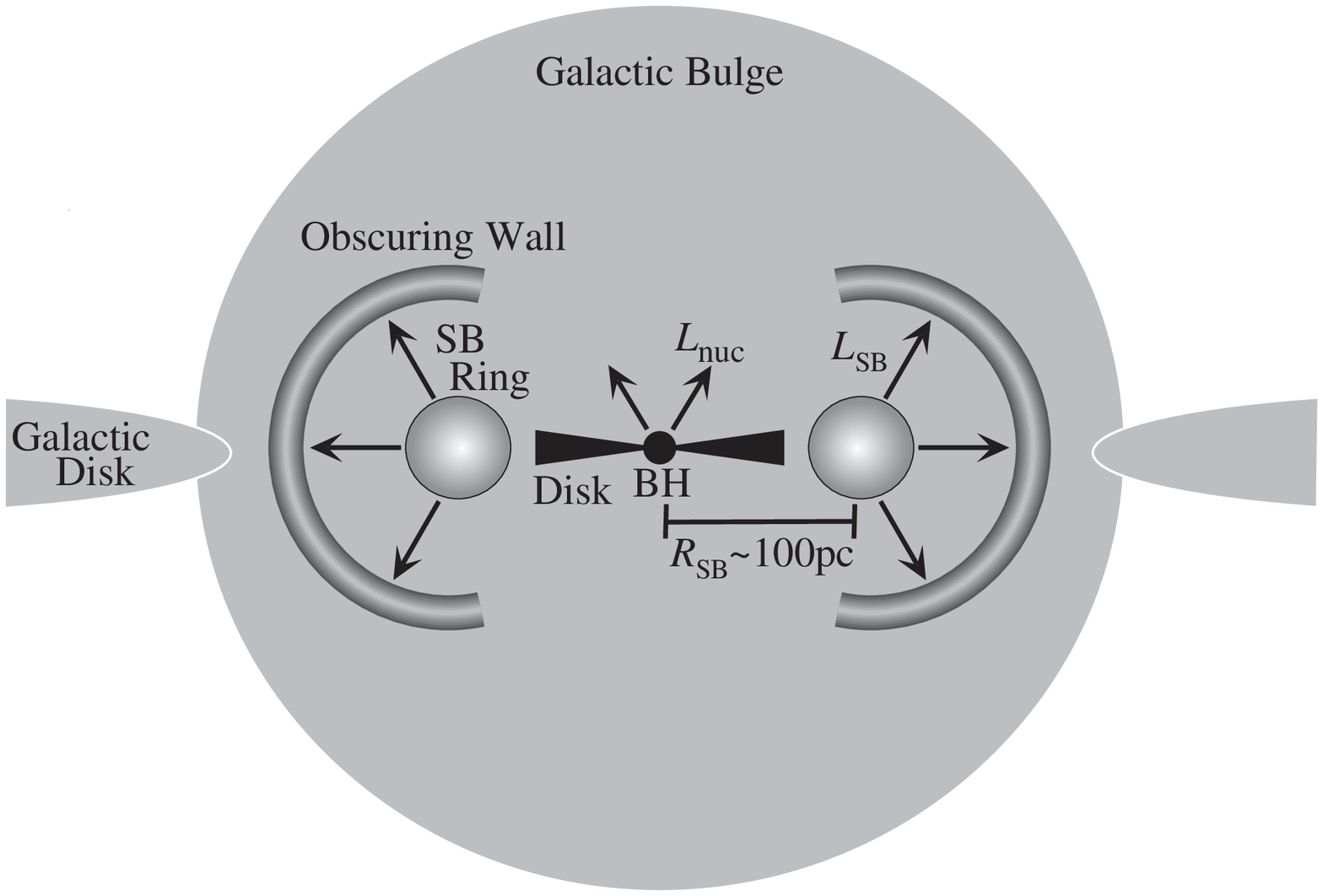}}
  \caption{
A schematic edge-on view of the circumnuclear structure 
in the present picture.
Here, we assume two components of radiation sources,
an AGN (luminosity $L_{\rm nuc}$) 
and a starburst ring (luminosity $L_{\rm SB}$).
The gravitational fields are modeled by four-component
gravity sources, i.e., a central black hole,
a gas disk, a starburst ring, and a galactic bulge.
The typical size of the starburst ring is around 100 pc,
so that an obscuring wall of several 100 pc is built up
due to the intensive radiation force by the starburst ring.
}
\end{figure}
\begin{figure}
  \centerline{\epsfxsize=400pt \epsfbox{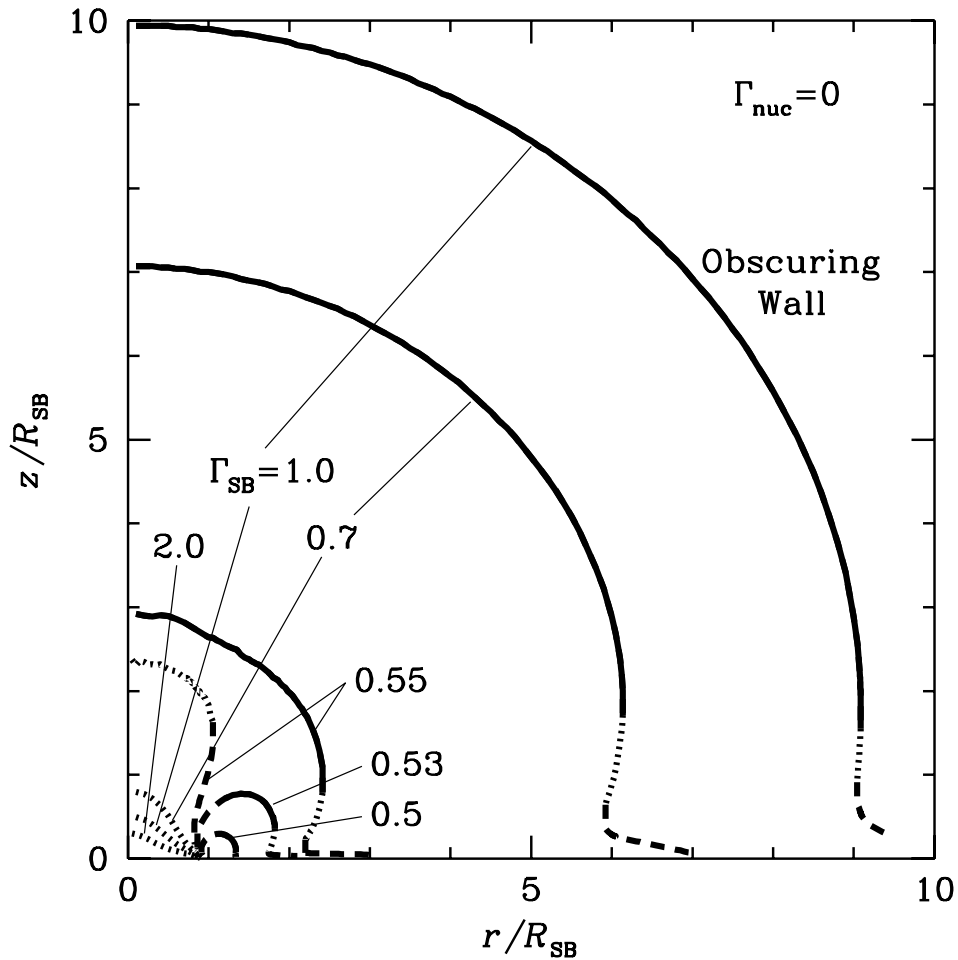}}
  \caption{
Equilibrium configuration of dusty gas on which 
the radiation force and gravity are exerted 
is shown in $r$-$z$ space for a wide variety of $\Gamma_{\rm SB}$.
The $r$ and $z$ are normalized by the starburst ring radius, $R_{\rm SB}$. 
Here, the nuclear luminosity is assumed to be null.
The solid and dashed curves represent vertically stable branches, while 
the dotted curved are vertically unstable branches.
Further, on the dashed curves 
the {\it radial} effective potential is locally maximal,
and therefore they are unstable points of saddle type, while
the solid curves are stable radially as well as vertically.
The solid curves show the final configuration of 
the stable obscuring wall.
}
\end{figure}

\begin{figure}
  \centerline{\epsfxsize=400pt \epsfbox{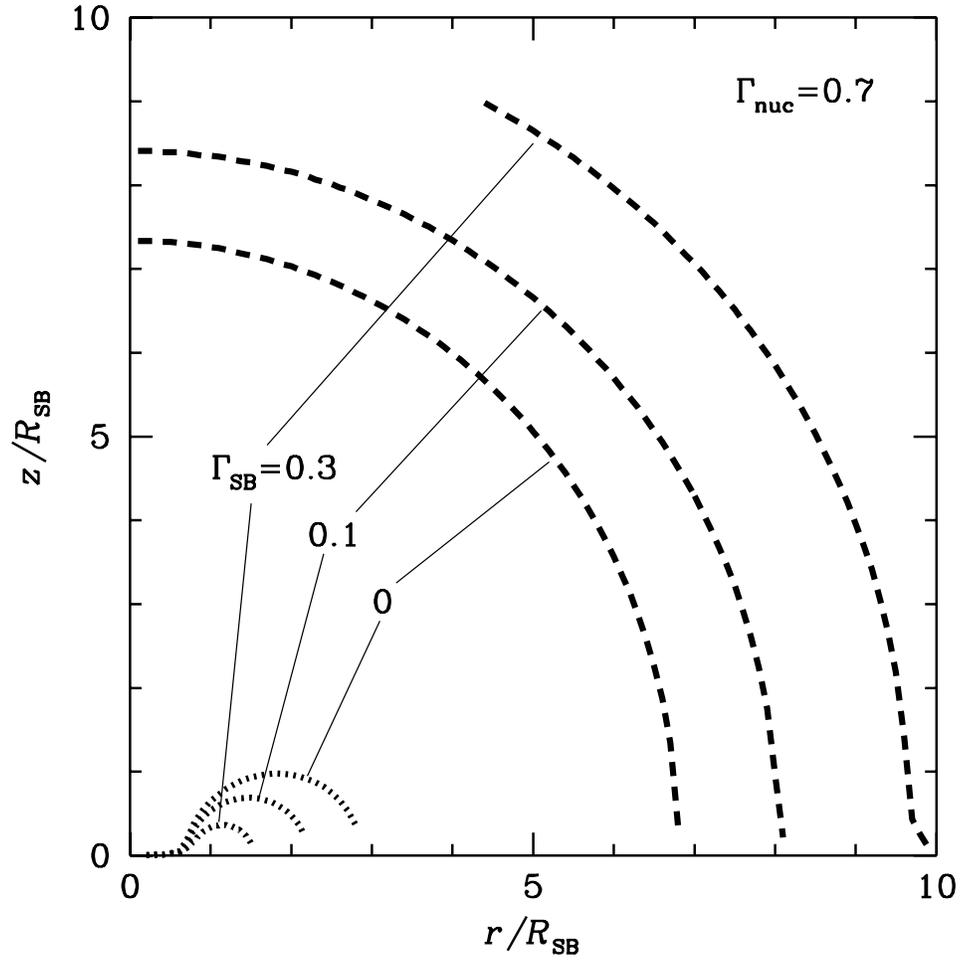}}
  \caption{
Same as Figure 2, but for a high nucleus luminosity 
($\Gamma_{\rm nuc} =0.7$).
In this case,
all the vertically stable branches do not satisfy
the radial equilibrium, regardless of starburst activity.
Hence, no stable wall forms.  
}
\end{figure}

\end{document}